\documentclass[12pt,preprint]{aastex}

\shorttitle{Variables in NGC 2099}
\shortauthors{Kang et al.}

\begin{document}
 
\title{Variable Stars in the Open Cluster NGC 2099 (M37)}

\author{Y. B. Kang}
\affil{Department of Astronomy and Space Science, Chungnam National University, Daejeon 305-764, Korea 
\\and Korea Astronomy and Space Science Institute, Daejeon 305-348, Korea}
\email{ybkang@cnu.ac.kr}
\author{S.-L. Kim}
\affil{Korea Astronomy and Space Science Institute, Daejeon 305-348, Korea}
\email{slkim@kasi.re.kr}
\author{S.-C. Rey}
\affil{Department of Astronomy and Space Science, Chungnam National University, Daejeon 305-764, Korea}
\email{screy@cnu.ac.kr}
\author{C.-U. Lee}
\affil{Korea Astronomy and Space Science Institute, Daejeon 305-348, Korea}
\author{Y. H. Kim}
\affil{Department of Astronomy and Space Science, Chungnam National University, Daejeon 305-764, Korea}
\author{J.-R. Koo}
\affil{Department of Astronomy and Space Science, Chungnam National University, Daejeon 305-764, Korea 
\\and Korea Astronomy and Space Science Institute, Daejeon 305-348, Korea}
\and
\author{Y.-B. Jeon}
\affil{Korea Astronomy and Space Science Institute, Daejeon 305-348, Korea}

\begin{abstract}
Time-series CCD photometric observations of the intermediate-age open cluster 
NGC 2099 were performed to search for variable stars. We also carried out 
$BV$ photometry to study physical properties of variables in the cluster. 
Using $V$-band time-series data, we carefully examined light variations of 
about 12,000 stars in the range of 10 $< V <$ 22 mag. 
A total of 24 variable stars have been identified; seven stars are previously 
known variables and 17 stars are newly identified. On the basis of 
observational properties such as light curve shape, period, and amplitude, 
we classified the new variable stars as nine $\delta$ Scuti-type pulsating 
stars, seven eclipsing binaries, and one peculiar variable star. 
Judging from the position of $\delta$ Scuti-type stars in the color-magnitude 
diagram, only two stars are likely to have the cluster membership. 
One new variable KV10 shows peculiar light variations with a $\delta$ 
Scuti-type short period of about 0.044 day as well as a long period of 
0.417 day. 
\end{abstract}

\keywords{open clusters and associations: individual (NGC 2099) --- 
delta Scuti --- binaries: eclipsing}

\section{Introduction}

Pulsating variable stars provide an ideal and important test bed for 
investigating the internal structure and evolution of stars; so-called 
asteroseismology. 
Open clusters have many benefits to investigate physical characteristics 
of Population I pulsating variables. 
Variable stars in a cluster can be restricted to their physical parameters 
by analysing the cluster with the same age, chemical abundance, distance, 
and interstellar reddening. 
Simultaneous photometry of hundreds of stars (both constant and variable) in a 
single CCD frame enables us to obtain accurate and effective time-series 
data. Furthermore, we can obtain high-precision photometric data for 
low-amplitude variables in open clusters since the crowding effect is not 
severe compared to the case of globular clusters. 
Consequently, the observations of open clusters are obviously important 
for the investigation on physical properties of small-amplitude variables 
such as $\delta$ Scuti-type pulsating stars \citep{kim01}. 
The open clusters with specific age range (e.g., intermediate-age 
of about 0.3 $\sim$ 1.0 Gyr) and at a reasonable distance of about 1 
$\sim$ 2 kpc, are the best observing target for the asteroseismological study 
of $\delta$ Scuti-type stars \citep{vis97}. 
Physical properties of $\delta$ Scuti-type stars in open clusters are 
reviewed in detail by \citet{rod01}. 

$\delta$ Scuti-type pulsating stars are main sequence or subgiant stars 
in lower part of the classical Cepheid instability strip with spectral types 
between A3 and F0. 
They have short pulsation periods of 0.02 $\sim$ 0.25 day \citep{bre00}. 
Generally, high amplitude $\delta$ Scuti-type stars with visual amplitude 
greater than 0.3 mag rotate slowly with $v \sin i$ less than 30 
km s$^{-1}$ and appear to be excited in single or double radial pulsating 
modes \citep{bre00}. 
However, most $\delta$ Scuti-type stars have small amplitudes less than 
0.1 mag and rotate fast with the average speed of about 150 km s$^{-1}$. 
Their light curves are very complicate to indicate the overlap of multiple 
pulsating periods; e.g. \citet{bre05} detected 79 frequencies for FG Vir. 
These multi-periodic $\delta$ Scuti-type stars excited in radial and 
non-radial modes are very important for asteroseismological tests on stellar 
structure and evolutionary models. 
\citet{gou96} suggested that multi-periodic $\delta$ Scuti-type stars near 
the zero-age main sequence are particularly suitable for determining the 
extent of the convective core and internal rotation rate. 
Asteroseismology on $\delta$ Scuti-type stars, however, has produced 
very few convincing results until now. 
Main reason for the lack of success is probably that the stars often rotate 
very fast to result in severe complexity of observing frequencies by 
rotational splitting of non-radial modes. 
The asteroseismological importance of observations for $\delta$ Scuti-type 
stars in open clusters was well described by \citet{kje00}. 

As our long term project for the survey of variable stars in open clusters, 
extensive time-series CCD observations have been performing for 
intermediate-age open clusters using an 1.0 m telescope at Mt. Lemmon Optical 
Astronomy Observatory (LOAO) in Arizona, USA. 
The primary goal of this project is to search for short-period (less than 
a few days) pulsating variables such as $\delta$ Scuti-type and $\gamma$ 
Doradus-type stars. 
It is also intriguing to examine possible relevance of characteristics of 
pulsating stars to the cluster parameters. 

In this paper, we present results of time-series CCD photometry for NGC 2099 
($\alpha_{2000}$ = 05$^h$52$^m$18\fs3, $\delta_{2000}$ = +32\arcdeg 33\arcmin 
10\farcs8).
The age of NGC 2099 about 400 Myr is relatively well-established putting the 
object among the intermediate-age open clusters from previous studies 
\citep[see][and references therein]{kal01, nil02}. 
It is a relatively rich and large cluster located at the low galactic latitude 
($b$ = 3\fdg1). \citet{kal01} and \citet{nil02} carried out deep CCD 
observations and obtained color-magnitude diagrams (CMDs) with well-defined 
main-sequence. However, there have been few systematic time-series 
observations for a large and representative sample of NGC 2099 so far. 
\citet{kis01} discovered only seven variables from their CCD observations. 

In Section 2, we present our observations and data analysis. Section 3 
describes physical properties of pulsating variables and binary stars 
detected from our observations. The details of a peculiar variable KV10 with 
complex light variations are reported in Section 4. We discuss the 
cluster membership of the variable stars in Section 5. A brief summary and 
conclusions are given in Section 6.

\section{Observations and data reductions}

Time-series observations of NGC 2099 were carried out on 12 nights  
in 2004 January, using a 2K CCD camera attached to the LOAO 1.0 m telescope 
in Arizona, USA. The Korea Astronomy and Space Science Institute (KASI) 
has been operating the telescope by remote control from Korea via a network 
connection. 
The field of view of a CCD image is about 22.2 $\times$ 22.2 arcmin$^{2}$ at 
the f/7.5 Cassegrain focus of the telescope. 
Total 581 images were obtained in the $V$ filter with 600 sec exposure time. 
Additional observations were performed for three nights in 2006 February 
in order to secure bright $\delta$ Scuti-type stars which are mostly 
saturated in long-exposure images. From this observing run, we obtained 639 
images in the $V$ filter with 10 sec exposure time. 
In order to minimize position-dependent external errors \citep{fra89}, 
we carefully controlled the telescope so that a star can be located at the 
same position in all CCD frames during our observing run. 
In order to construct the CMD of NGC 2099, $BV$ photometric 
observations were also made for two nights in 2004 October and one night in 
2006 February.  Typical seeing (FWHM) was about 2.2 arcsec. Figure 1 shows an 
observing field of the NGC 2099.

\begin{figure}
\epsscale{1.0}
\plotone{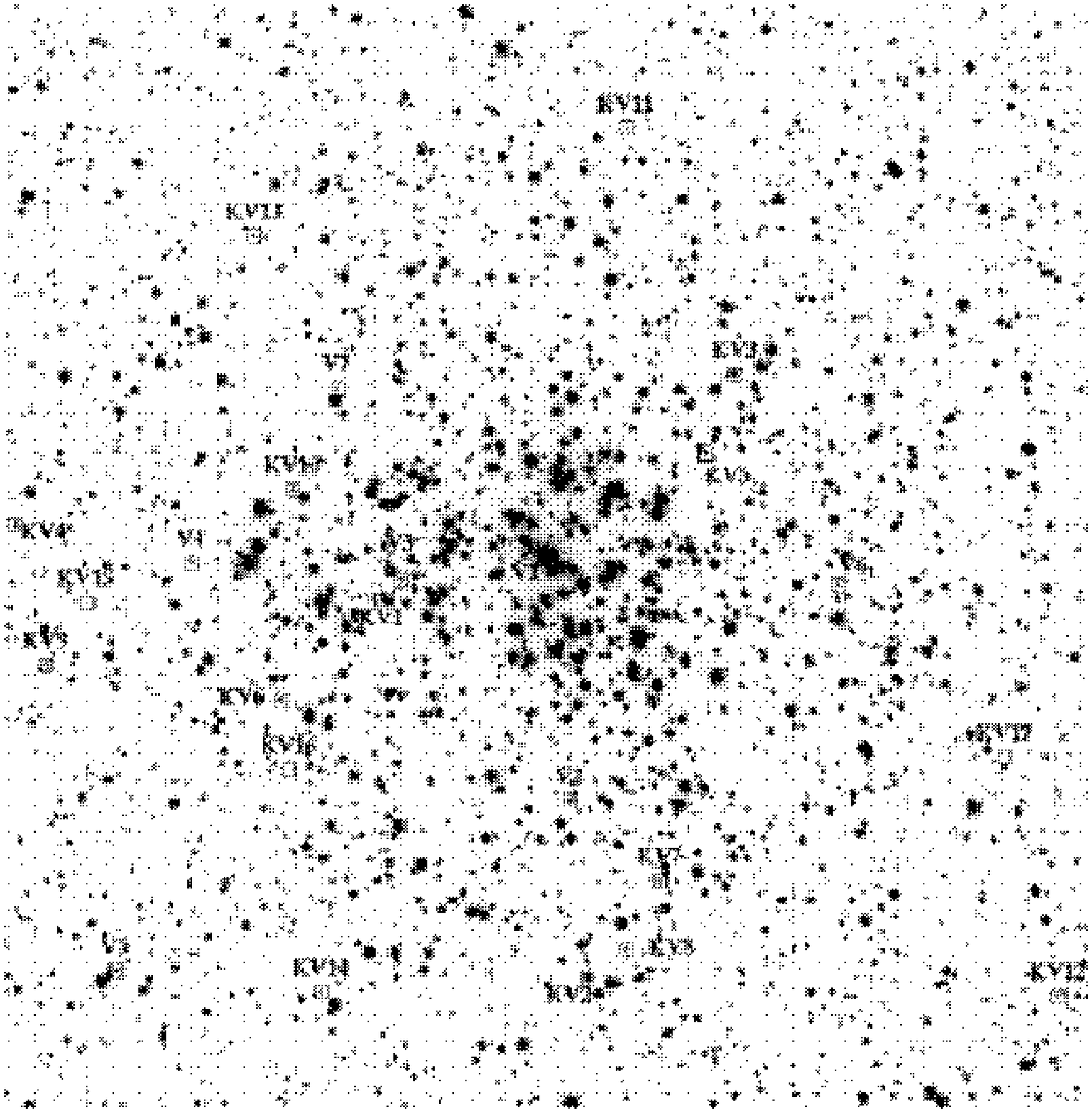}
\caption{Observed CCD field (22.2 \ensuremath{\times} 22.2 arcmin$^{2}$) 
of the open cluster NGC 2099. Identification of variable 
stars are also marked. North is up and east is to the left.\label{fig1}}
\end{figure}

\placefigure{fig1}

Instrumental signatures of each frame were removed and calibrated using 
the bias, dark, and flat field frames, with aid of the IRAF/CCDRED package. 
We obtained instrumental magnitudes of stars from the empirical point spread 
function (PSF) fitting method in the IRAF/DAOPHOT package \citep{mas92}. 

Since we did not make observations for standard stars, we have tied our 
data to the $BV$ CCD photometric secondary standards in NGC 2099 from the 
data of \citet{kal01}. We compared our CCD photometry with the data from 
\citet{kal01}. 
Our $B$ \& $V$ magnitudes are in good agreement with those of 
\citet{kal01}: $\Delta V$ = $-$0.022 and $\Delta B$ = $-$0.032.

In order to normalize instrumental magnitudes of our time-series $V$ frames, 
we applied an ensemble normalization technique \citep{gil88}. 
We selected a few tens of bright and unsaturated normalization stars 
with 15 $< V <$ 17 mag, which are non-variable stars and are 
not located at the edge of CCD frame. Coefficients of the following 
equations are calculated for each CCD frame ;

\begin{center}
$B=b+a_{1}+a_{2}(B-V)+a_{3}X+a_{4}Y$,
\end{center}
\begin{center}
$V=v+c_{1}+c_{2}(B-V)+c_{3}X+c_{4}Y$,
\end{center}

\noindent
where $B$ \& $V$ are standard magnitudes and $b$ \& $v$ are instrumental 
magnitudes of time-series images for normalization stars respectively. 
$X$ and $Y$ are coordinates of the normalization stars in a CCD frame. 
Using the above equations, we normalized all stars for each 
CCD frame. With this technique, we corrected color (a$_{2}$, c$_{2}$) and 
position (a$_{3}$, c$_{3}$ and a$_{4}$, c$_{4}$) dependent variations of the 
observation system and/or atmospheric differential extinction for all 
time-series CCD frames \citep{kim01}.

\section{Physical properties of variable stars}

We examined light variations of about 12,000 stars by visual inspection. 
Excluding saturated stars brighter than $V$ = 13 and 10 mag in the long 
and short exposure images, respectively, we identified 24 variable stars in 
the observing field. Among these, 17 (KV1 - KV17) variables were newly 
discovered by our observations. 
A finding chart of variable stars is shown in Figure 1. Basic properties 
including mean magnitude $<$$V$$>$, mean color $<$\bv$>$, and variable types 
of variable stars are presented in Table 1. 

\placetable{tabl1}

We derived the periods of variable stars using the discrete Fourier analysis 
\citep{sca89,kim01}. Especially, for all stars in the $\delta$ Scuti 
instability strip, we investigated the signal to noise amplitude ratio 
(S/N) of a frequency with the biggest power in a power spectrum of each star, 
in order to search for variability. The S/N ratio greater than 4.0 
\citep{bre93} has been used as a detection criterion of pulsating frequencies; 
for example, analysis of the $\delta$ Scuti-type star 57 Tau by \citet{pap00}.

\subsection{Previously known variable stars}

\citet{kis01} discovered seven (V1 - V7) variable stars in the field of 
NGC 2099 through the time-series measurements with $R_C$ filter. 
Table 2 compares the basic properties of variables obtained from our data with 
those of \citet{kis01}. 
They suggested that two (V1 and V2) stars are long period eclipsing binary 
candidates without any definite period determinations. Three (V3, V4, and V7) 
of them were considered to be W UMa-type eclipsing binaries.  The other two 
(V5 and V6) stars were suggested to be high-amplitude $\delta$ Scuti-type stars. 

\placetable{tabl2}

From the detection of only one deep minimum in their light curves 
of V1 and V2, \citet{kis01} classified these stars as long period binaries. 
However, we could not determine period and epoch of minimum, since most of 
their observations showed no significant light variation. 
As shown in the upper panels of Figure 2, any distinct feature of light 
variations of V1 and V2 were not found. Due to our limited data points 
obtained from short observing run ($\sim$ 10 hours), we failed to detect any 
fading and to confirm the characteristics of these two stars. 
More time-series observations with long observing run should be made in 
order to clarify the properties of these stars.

\begin{figure}
\epsscale{.80}
\plotone{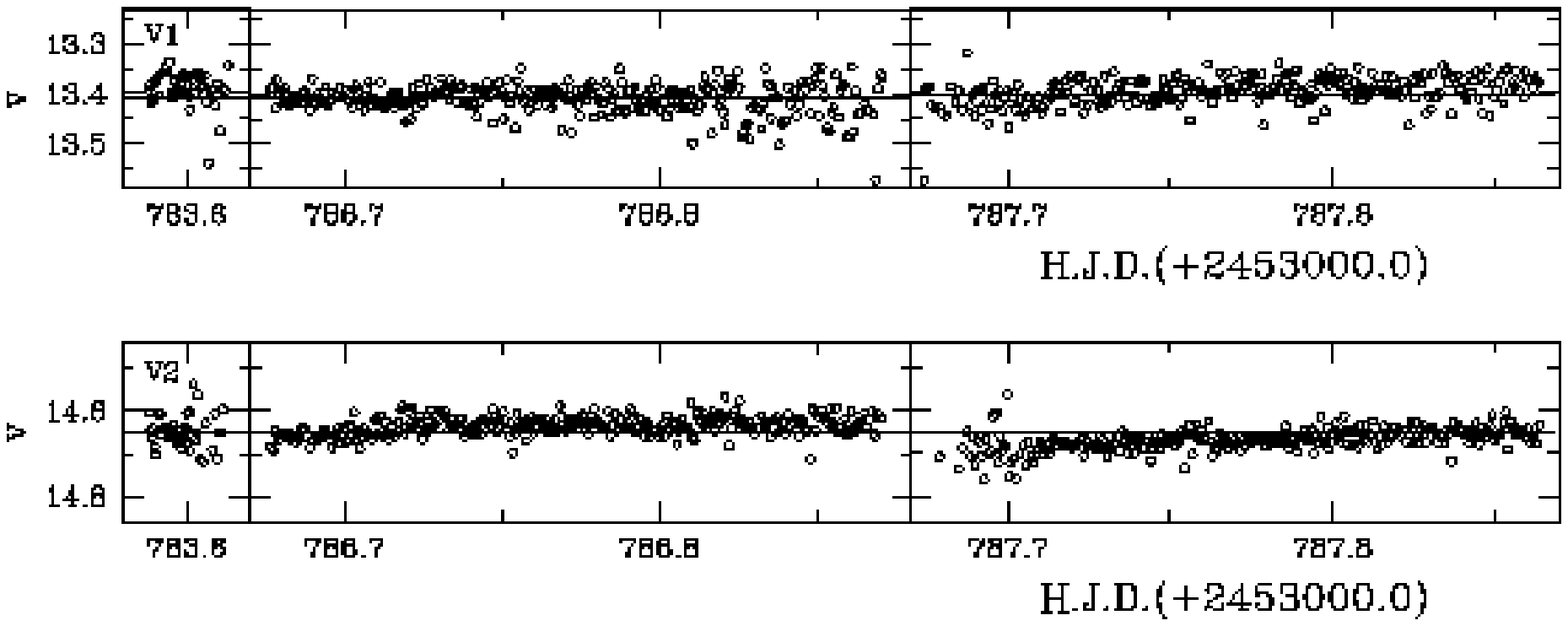}
\plotone{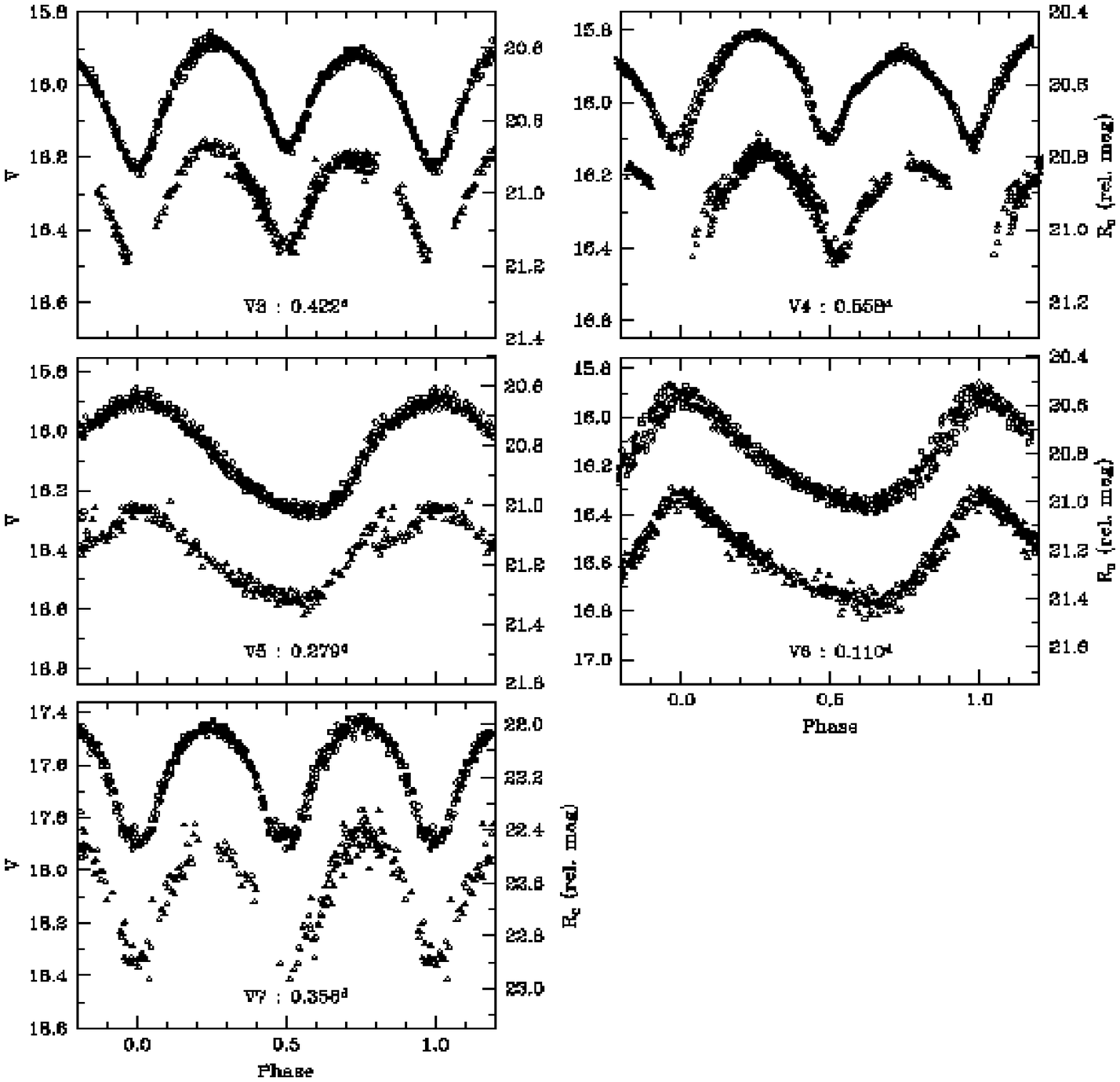}
\caption{($Upper$ $panels$) Light variations of V1 and V2. We could not detect 
any fading for these two Algol-type eclipsing binaries discovered by 
\citet{kis01}. ($Lower$ $panels$) Phase diagrams of five (V3 - V7) variables. 
Open circles and triangles are our $V$ band data and $R_C$ band data of 
\citet{kis01}, respectively.
\label{fig2}}
\end{figure}

\placefigure{fig2}

Lower panels of Figure 2 show the overall shape of our phase diagrams for 
other five variables which are in good agreements with those of 
\citet{kis01}. 
Especially, we could determine more accurate periods and epochs for all 
variables since we have much more data points which fully cover the phase 
diagrams. 
The light curves of V3, V4, and V7 show typical shapes of W UMa-type binary. 
Especially, our data for V7 show remarkably clear and complete light curve 
compared to the previous one of \citet{kis01}, which allows us to verify this 
star as a typical W UMa-type binary. 
We confirmed that V3 and V4 show slightly different brightness of maxima 
between the orbital phases of 0.25 and of 0.75, which gives us a hint for 
the existence of spots on the stellar surface \citep{wil94}. 

\citet{kis01} suggested that V5 might be high-amplitude $\delta$ Scuti-type 
star or field RRc-type star. However, our light curve shows that V5 has a 
typical shape of RRc-type star rather than that of high amplitude $\delta$ 
Scuti-type star. 
Our light curves of V6 together with \citet{kis01} definitely show a 
characteristic of high-amplitude $\delta$ Scuti-type star. 
We applied the multi-frequency analysis \citep{kim96} to the data of V5 and 
V6. V5 does not show any significant peak on the frequency spectra after 
prewhitening the main frequency of $f_1$ = 3.588 c/d (cycles per day). 
For the variable V6, we found two frequencies of $f_1$ = 9.104 c/d with 
amplitude of $A_1$ = 204.2$\pm$1.4 mmag (S/N = 37.2) and $f_2$ = 15.016 c/d 
with amplitude of $A_2$ = 32.7$\pm$1.4 mmag (S/N = 14.1). 
The frequency ratio of $f_1$/$f_2$ = 0.606 is very similar to that of 
theoretical radial modes $P_2/P_0$ = 0.616 \citep{bre79}, indicating that 
$f_1$ is a fundamental radial mode (F-mode; $P_0$) and $f_2$ is a 2nd 
overtone radial mode (2H-mode; $P_2$). These indicate that V6 is a high 
amplitude $\delta$ Scuti-type star with double radial modes of $P_0$ and 
$P_2$. The larger dispersion shown in the phase diagram of V6 is caused by 
the double-mode nature of pulsation.

\subsection{New pulsating variable stars}

Among 17 new variable stars discovered by our observations, nine stars are 
identified as pulsating variables with characteristics of $\delta$ Scuti-type 
star (see Table 1 for identification). In Table 3, we list the pulsating 
frequencies of new variable stars derived from the multiple frequency analysis 
\citep{kim96}.
Their power spectra are shown in Figures 3 and 4. Light variations and phase 
diagrams of nine $\delta$ Scuti-type stars are shown in Figures 5 and 6. 
The solid lines in Figure 5 represent synthetic curves computed from our 
multiple frequency analysis.

\begin{figure}
\epsscale{.60}
\plotone{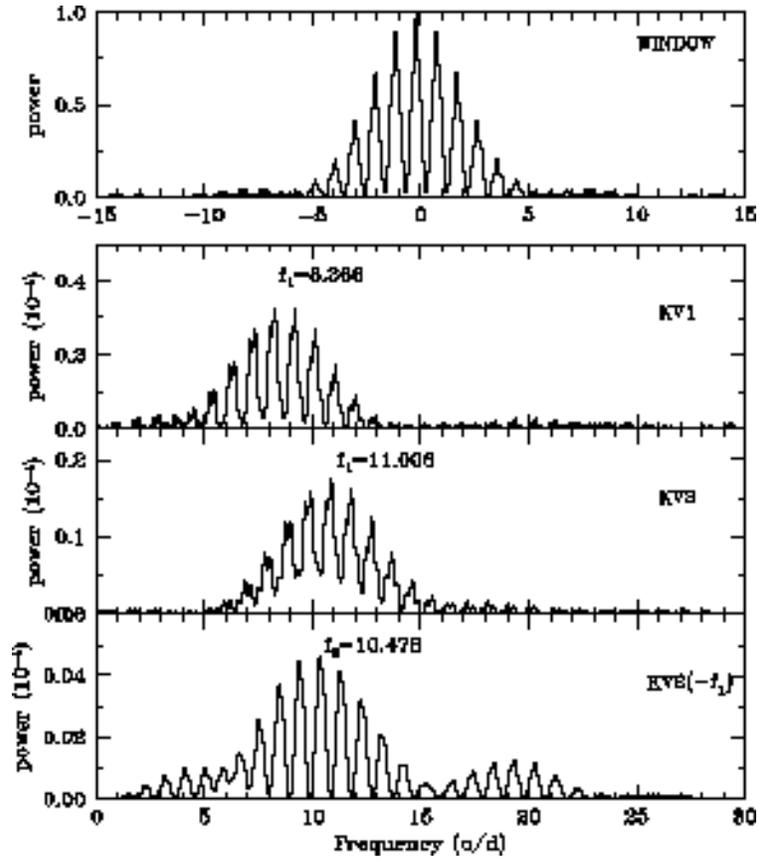}
\caption{Power spectra of $\delta$ Scuti-type stars KV1 and KV2. A window 
spectrum is shown in the top panel.\label{fig3}}
\end{figure}

\begin{figure}
\epsscale{.70}
\plotone{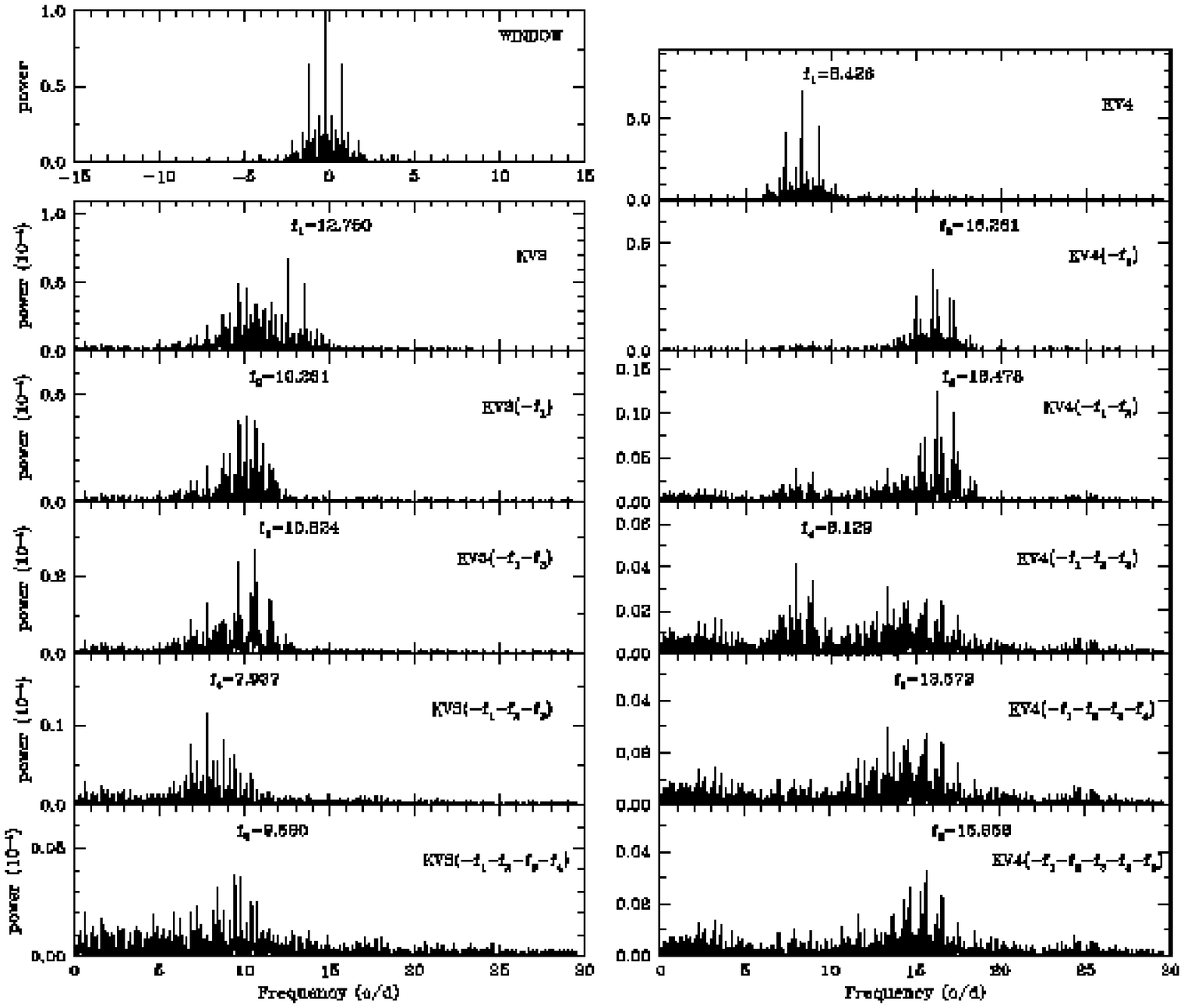}
\plotone{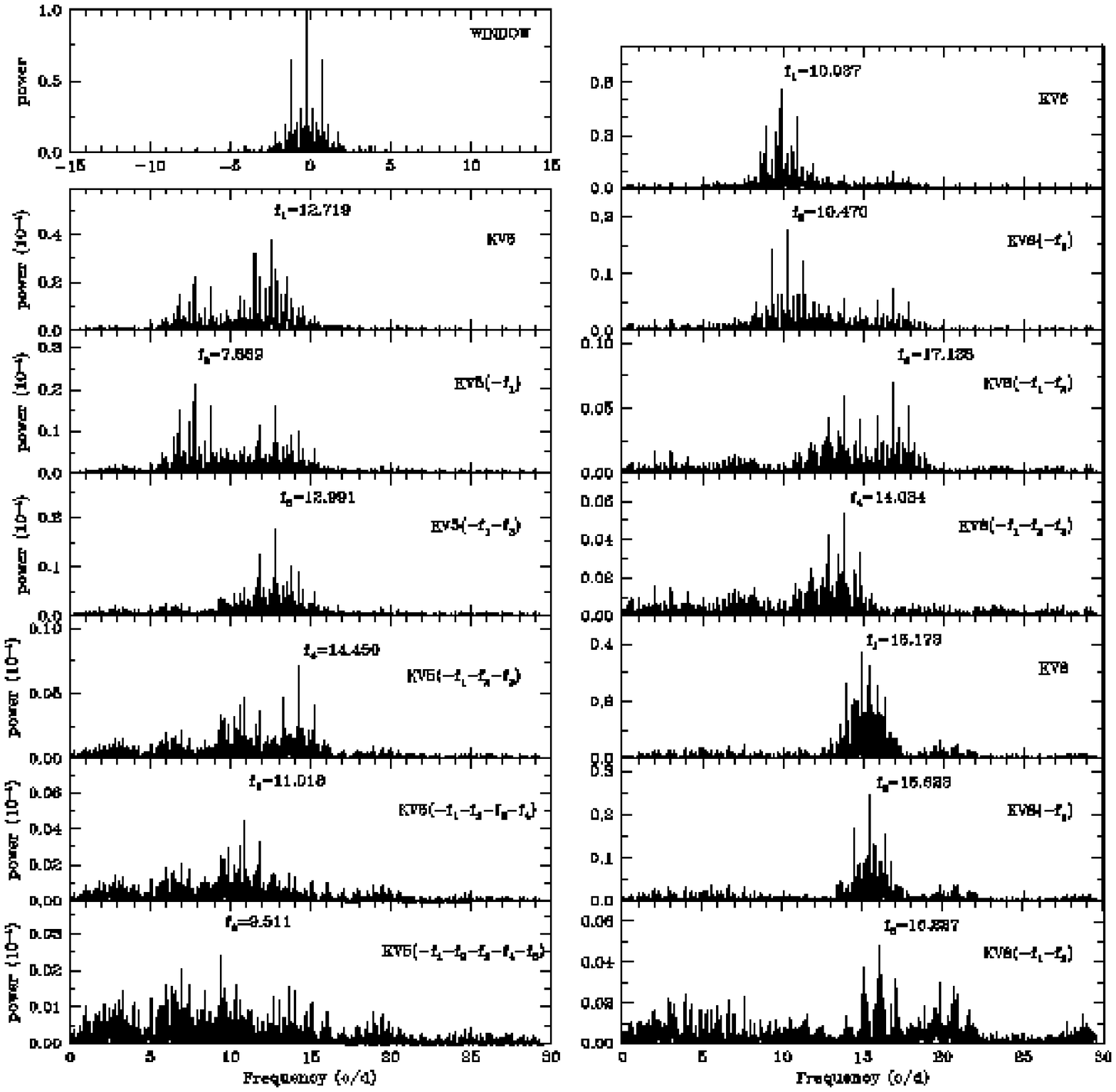}
\caption{Power spectra of five (KV3 - KV6, and KV8) $\delta$ Scuti-type stars. 
Window spectra are shown in the top left.\label{fig4}}
\end{figure}

\begin{figure}
\epsscale{.50}
\plotone{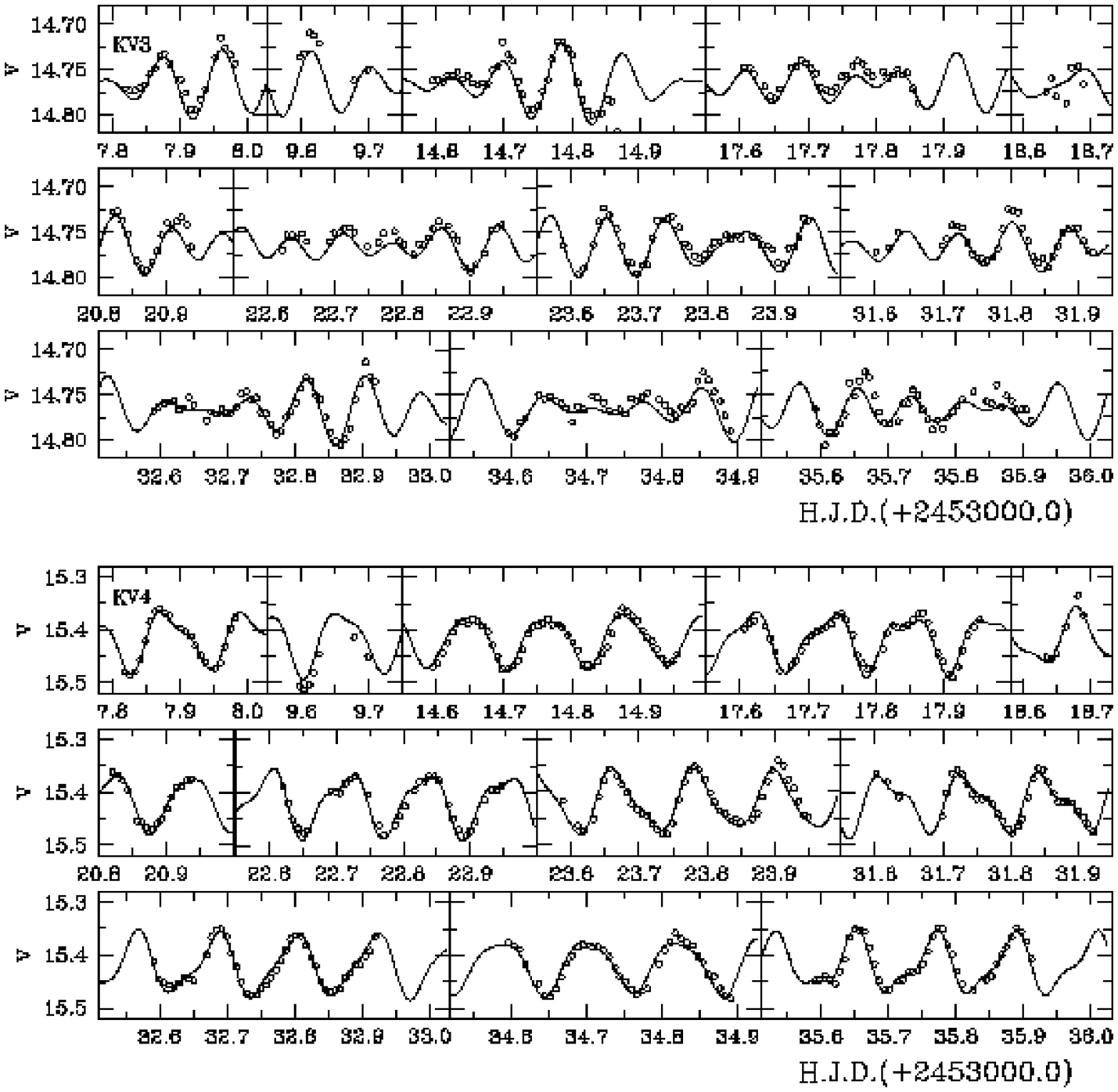}
\plotone{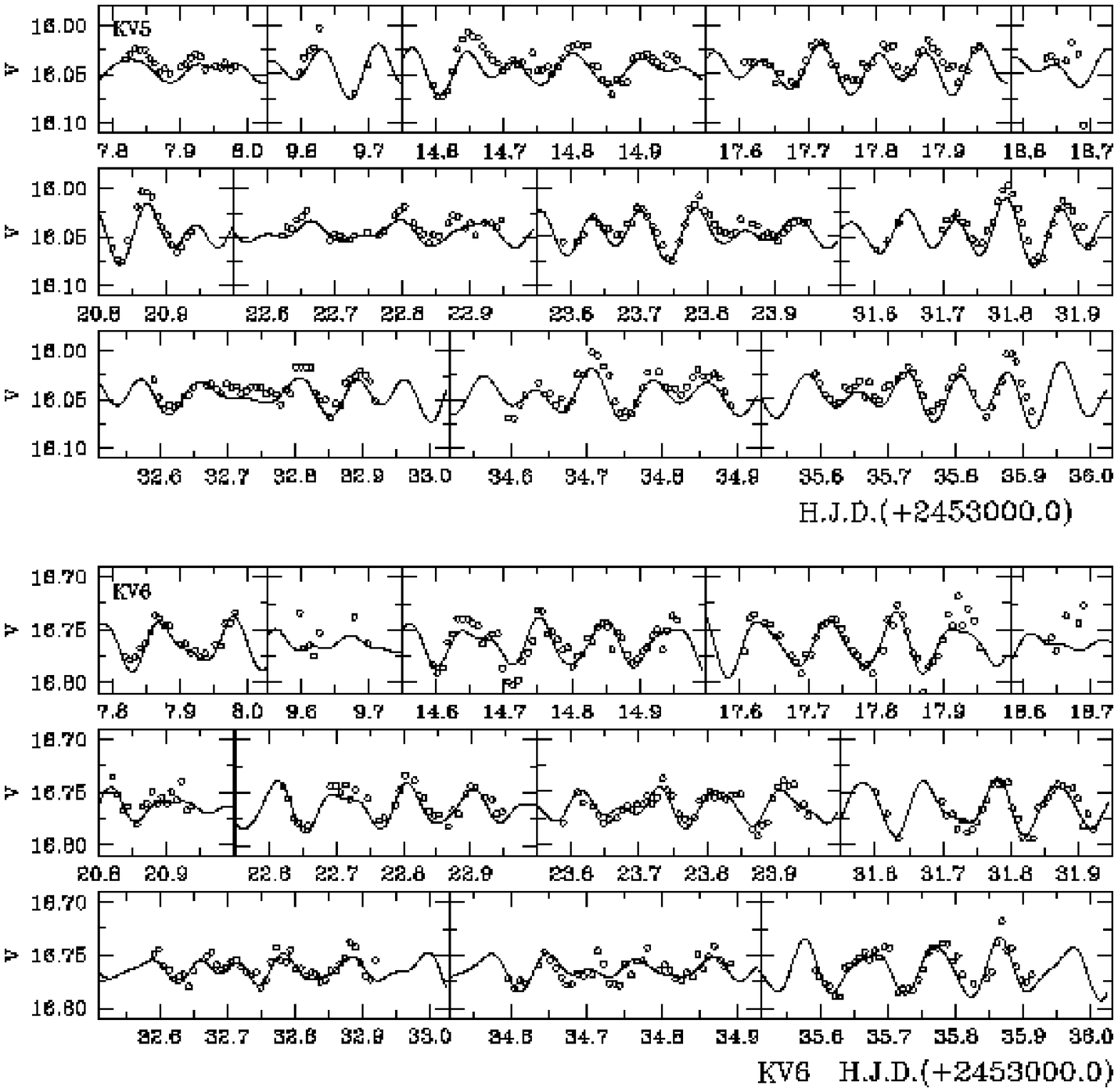}
\plotone{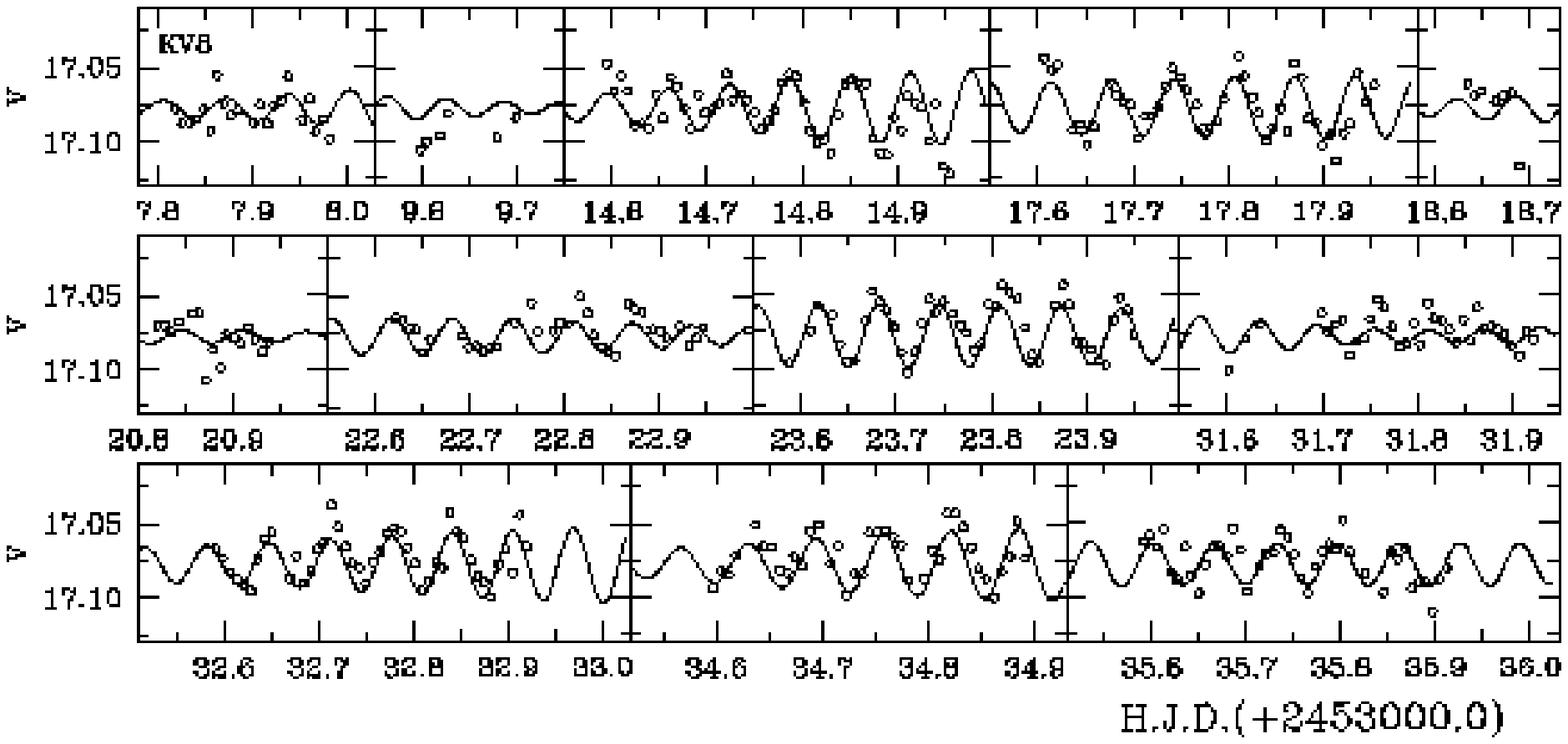}
\caption{Light variations of five (KV3 - KV6, and KV8) $\delta$ Scuti-type 
stars discovered from our observations. Synthetic curves computed 
from our multiple frequency analysis are also superimposed.\label{fig5}}
\end{figure}

\begin{figure}
\epsscale{.80}
\plotone{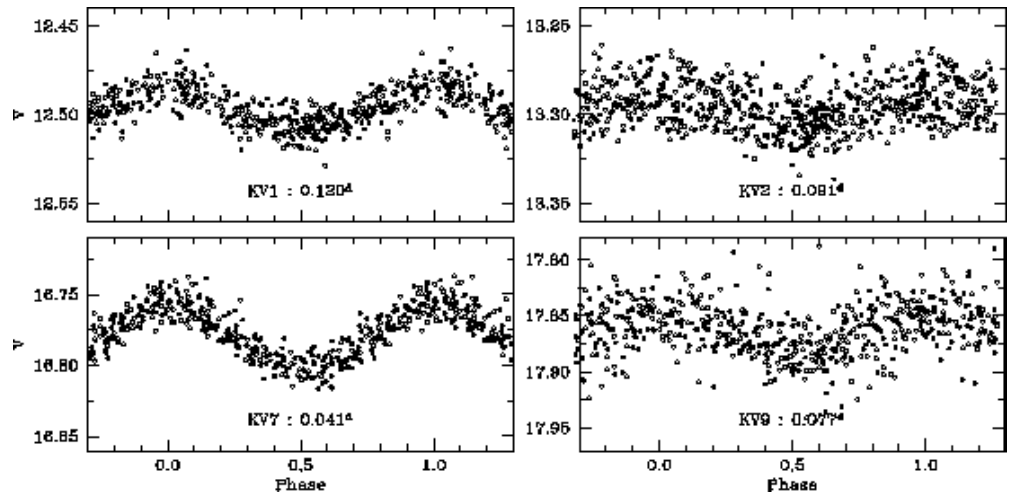}
\caption{Phase diagrams for two (KV1 and KV7) $\delta$ Scuti-type stars 
and two (KV2 and KV9) $\delta$ Scuti-type candidates.
Large scatters of KV2 might be originated from double frequencies as shown 
in Figure 3.\label{fig6}}
\end{figure}

\placetable{tabl3}
\placefigure{fig3}
\placefigure{fig4}
\placefigure{fig5}
\placefigure{fig6}

The data of KV1 and KV2 are obtained from only short exposure observations due 
to their saturation in long exposure images. They have only one and two 
frequencies, respectively, with small amplitudes. 
Although we have short time span of the short exposure observations for these 
stars, we suggest that they are $\delta$ Scuti-type star (KV1) and $\delta$ 
Scuti-type star candidate (KV2). 
Five (KV3, KV4, KV5, KV6, and KV8) variable stars have more than three 
frequencies. 
They show frequencies in the range of 7.937 c/d $\sim$ 16.478 c/d which 
are typical values for the $\delta$ Scuti-type stars. 
They have closely-separated frequencies (frequency ratio around 1.0: 
$f_2$/$f_3$ = 0.948 for KV3, $f_1$/$f_4$ = 1.037 for KV4, 
$f_1$/$f_3$ = 0.979 for KV5, $f_1$/$f_2$ = 0.959 for KV6, 
and $f_1$/$f_2$ = 0.971 for KV8), which indicate the excitation of non-radial 
modes \citep{bre79}. In the case of KV7, we detected only one frequency of 
f$_1$ = 24.308 c/d with large power, which is the typical frequency for the 
$\delta$ Scuti-type stars. Although KV9 shows some scatter on the phase 
diagram, we detected one dominant frequency of f$_1$ = 13.020 c/d with 
large S/N ratio of 8.0. Thus we suggest that KV9 is a $\delta$ Scuti-type 
star candidate.

\subsection{New eclipsing binaries}

We discovered seven new eclipsing binaries, which show clear characteristics 
of W UMa-type binary except for KV12 (see Table 1 for the identification). 
In Figure 7, we show the phase diagrams for these newly discovered eclipsing 
binaries. We determined their orbital periods using the phase-match technique 
\citep{hof85}. 
KV11 shows the shallow secondary maximum (around orbital phase 0.75) and 
a hump just after the secondary eclipse, which is possibly due to the 
existence of surface inhomogeneity, i.e. effect of the spot. Although definite 
type of KV12 is ambiguous due to its incomplete light curve, it also appears 
to be an eclipsing binary. 
KV13 and KV15 show the most clear and typical W UMa-type light curves, with 
slightly different brightness of maxima probably due to the existence of a 
spot on the stellar surface. Although there are some scatters over the whole 
phase of light curves of KV14, KV16, and KV17, they also appear to be 
W UMa-type binaries. In Table 4, we summarize light curve parameters 
(period, epoch of minimum, and amplitude) of these eclipsing binaries. 

\begin{figure}
\epsscale{.80}
\plotone{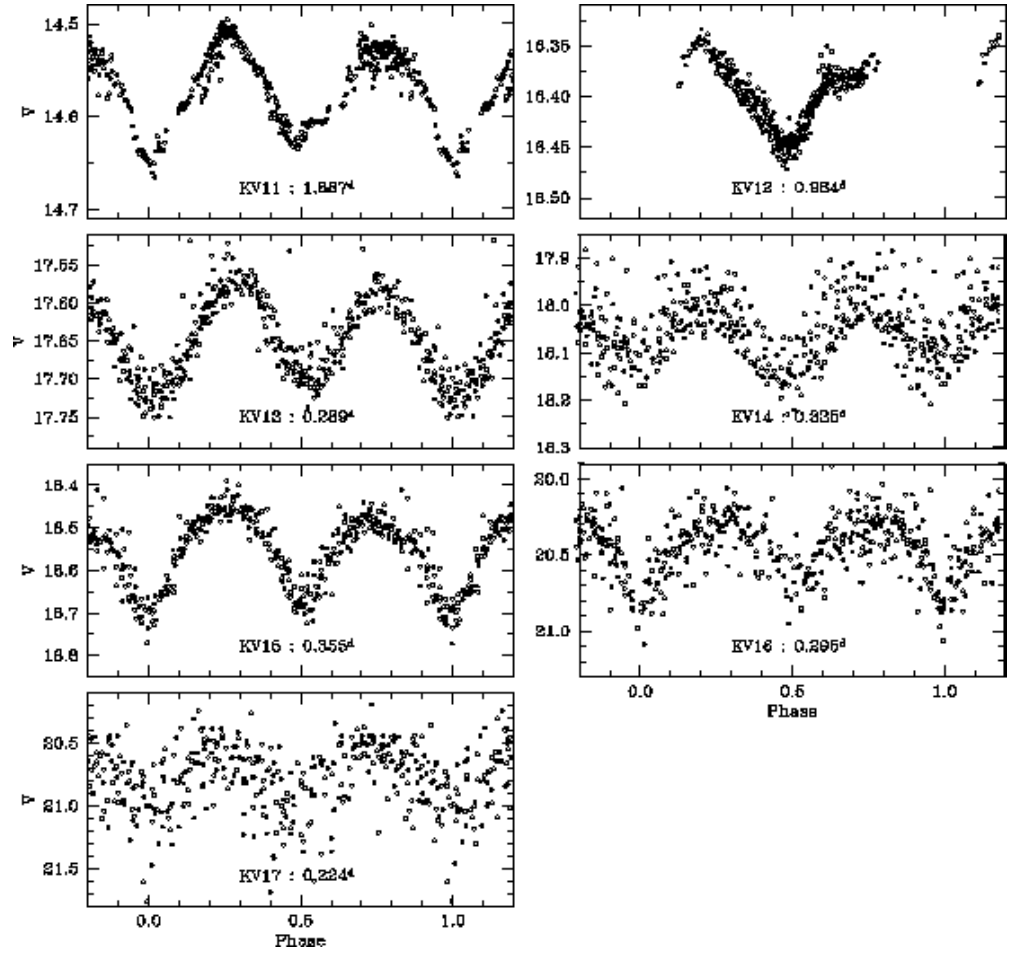}
\caption{Phase diagrams of newly discovered seven (KV11 - KV17) eclipsing binary stars.\label{fig7}}
\end{figure}

\placefigure{fig7}
\placetable{tabl4}

\section{Peculiar light variations of KV10}

Among the new variable stars discovered from our observations, KV10 is the most 
interesting object with complex light variations. From our multiple 
frequency analysis \citep{kim96}, we found that KV10 has two frequencies 
with similar amplitudes: $f_1$ = 22.796 c/d (high) with 32.9 mmag and 
$f_2$ = 2.396 c/d (low) with 24.2 mmag. 
Figure 8 displays the complex light variations of KV10. Upper panel shows 
original light variations. Lower panels are phase diagrams of residuals 
obtained from the subtraction of $f_1$ to the data, assuming two different 
periods of 1/$f_2$ = 0.417 day or two times of it.

\begin{figure}
\epsscale{.80}
\plotone{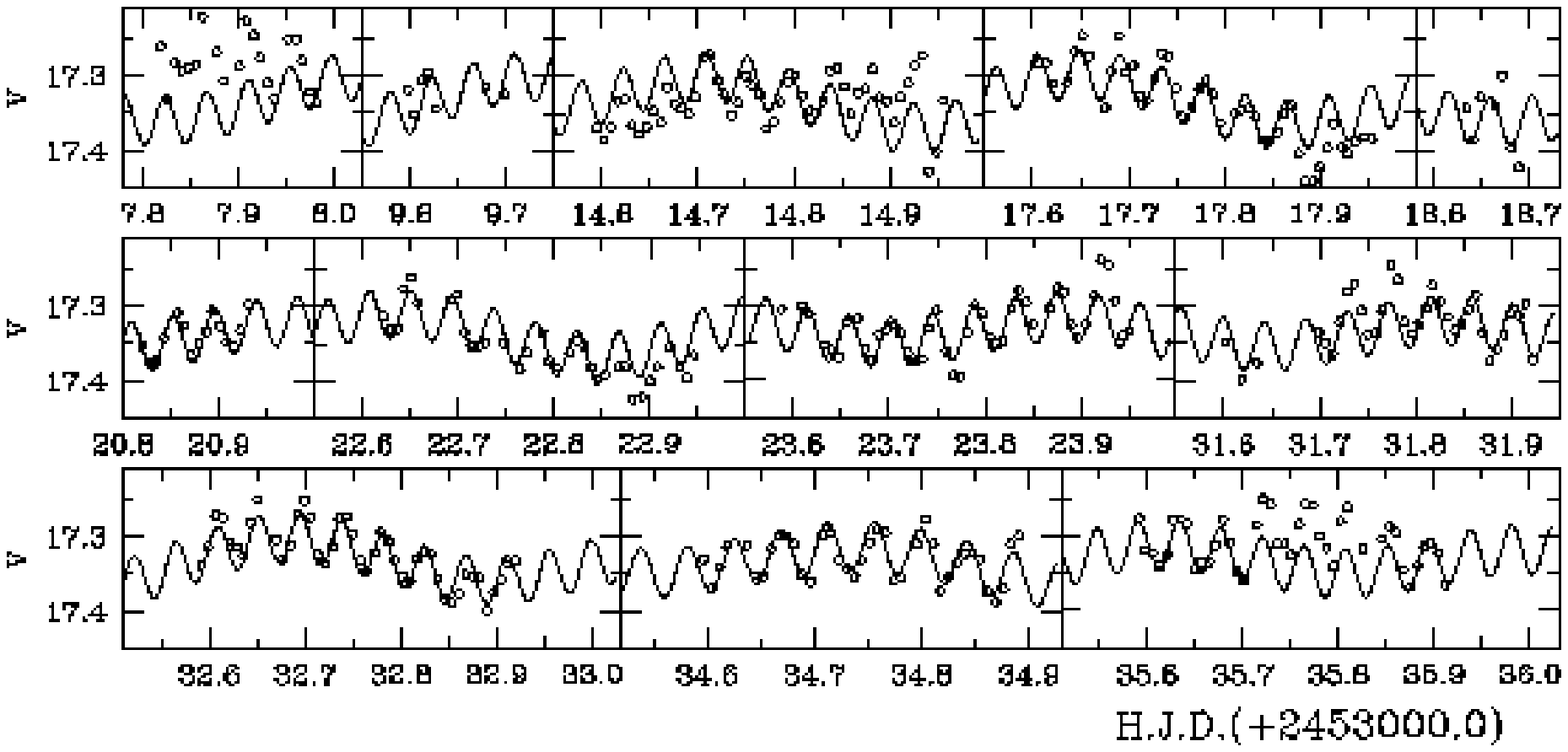}
\plotone{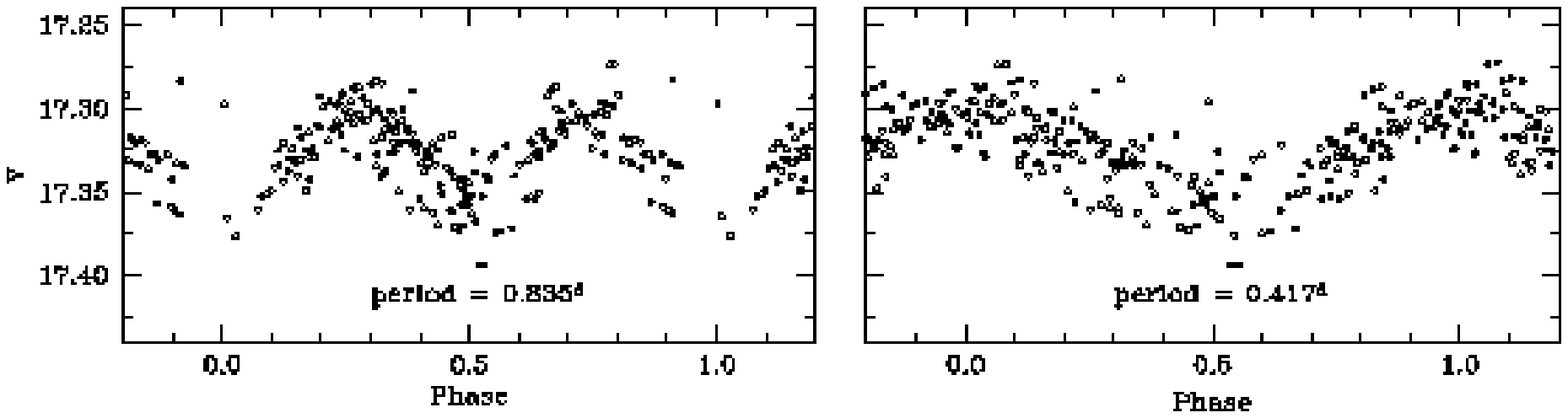}
\caption{Light variations of the peculiar variable KV10. ($Upper$ $panel$) 
Original time-series data. The solid line is synthetic curve computed from 
our multiple frequency analysis with $f_1$ = 22.796 c/d and $f_2$= 2.396 c/d. 
($Lower$ $panels$) Residuals obtained from the subtraction of $f_1$ to the 
data. We present two different-type light variations by assuming different 
periods, i.e. ellipsoidal variations with period of 0.835 day 
($left$ $lower$ $panel$) and pulsating variable with a period of 0.417 day 
($right$ $lower$ $panel$).
\label{fig8}}
\end{figure}

\placefigure{fig8}

There might be two possible interpretations for these peculiar light 
variations of KV10. 
First, KV10 can be a hybrid pulsator to be excited in both $\gamma$ 
Doradus-type g-mode with a period of 0.417 day and $\delta$ Scuti-type 
p-mode with a period of 0.044 day. This kind of hybrid pulsation has been 
known for two stars of HD 8801 \citep{hen05} and HD 209295 \citep{han02}.
Second, KV10 may be a binary system with an orbital period of 0.835 day with 
ellipsoidal variations and one of the components in this binary system is a 
$\delta$ Scuti-type pulsator. If the binary system has a semi-detached 
configuration, one of the components fills its Roche lobe and then makes 
ellipsoidal variations with its rotation. For example, from the photometric 
and spectroscopic observations of a triple system HD 207651, \citet{hen04} 
found two frequencies of 15.45 c/d resulted from $\delta$ Scuti-type 
pulsation and 1.36 c/d from ellipsoidal variation. 
On the other hand, a few tens of $\delta$ Scuti-type pulsating components 
in semi-detached eclipsing binary systems have been discovered
\citep[see the recent catalogue by][and references therein]{soy06}.

Judging from the available near-infrared color indices from the 2MASS data 
(from the SIMBAD\footnote{The SIMBAD database, operate at CDS, Strasbourg, 
France}), ($J-H$) = 0.448$\pm$0.135 and ($J-K$) = 0.451$\pm$0.157, 
KV10 seems to be late G or early K type star \citep[see][Table II]{bes88}. 
This indicates that KV10 may not be an example of hybrid pulsator 
with A - F spectral type exhibiting both $\delta$ Scuti-type and $\gamma$ 
Doradus-type pulsations. 
Therefore, we prefer the second interpretation, i.e. KV10 is a semi-detached 
binary system with an orbital period of 0.835 day whose primary component is 
a $\delta$ Scuti-type pulsator and secondary one is a late-type giant to fill 
its Roche lobe inducing ellipsoidal variations. 

\section{Cluster membership of the variable stars}

The observed CMD of NGC 2099 is shown in Figure 9. The thick and 
thin solid lines represent adopted empirical zero-age main sequence (ZAMS) 
from \citet{sun99} and theoretical isochrone with log t = 8.65 and 
Z = 0.019  \citep{gir00}, respectively. 
The best fit of the ZAMS and isochrone to the observed CMD gives the 
interstellar reddening of E($B-V$) = 0.21 and the distance modulus of 
($V-M_{V}$)$_{0}$ = 11.4. These values are in good agreements 
with the previous results by \citet{kal01}; E($B-V$) = 0.21 
and ($V-M_{V}$)$_{0}$ = 11.55.

\begin{figure}
\epsscale{1.0}
\plotone{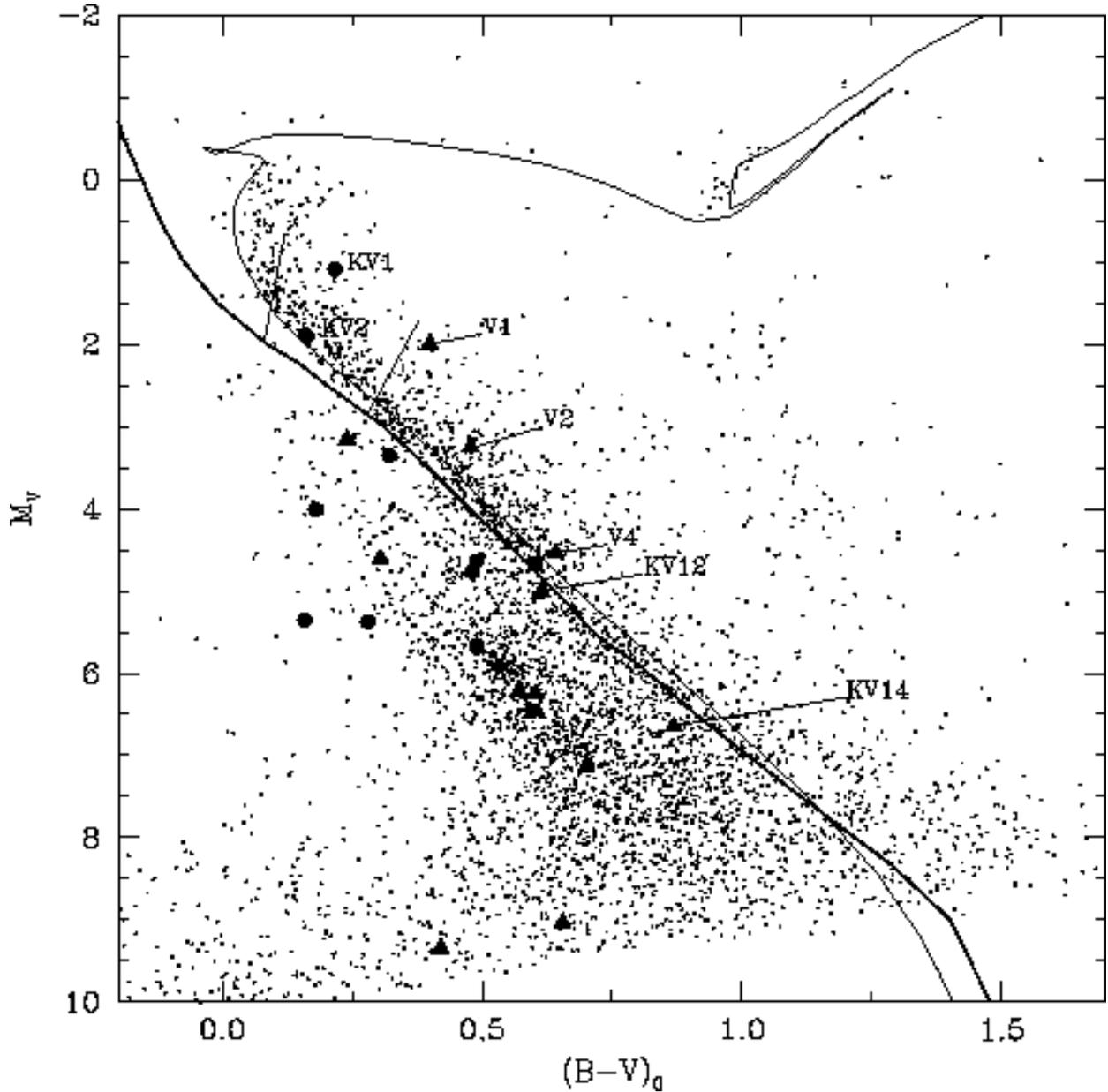}
\caption{Positions of 24 variable stars in the color-magnitude diagram of 
NGC 2099. Different symbols indicate different type of variables; filled 
circles for $\delta$ Scuti-type pulsating stars, filled triangles for 
eclipsing binaries, asterisk for peculiar variable KV10, and filled square 
for V5 as RRc-type star. 
The thick and thin solid lines represent the adopted empirical zero-age main 
sequence from \citet{sun99} and theoretical isochrone from \citet{gir00}, 
respectively. Solid bars, nearly perpendicular to the ZAMS, represent 
$\delta$ Scuti instability strip \citep{bre79}.\label{fig9}}
\end{figure}

\placefigure{fig9}

In Figure 9, we show the position of 24 variable stars. 
Among 224 stars within the $\delta$ Scuti instability strip, only two 
(KV1 and KV2) $\delta$ Scuti-type stars are detected. These two 
$\delta$ Scuti-type stars have the lowest amplitudes of about 
$\Delta$$V$ $\sim$ 0.02 mag among our newly discovered variables. 
The membership probability of KV1 based on the proper motion data by 
\citet{zha85} indicates 0.90, which confirms KV1 as a possible member star 
in the cluster. 
Assuming that most of the stars in the instability strip are cluster members, 
the fraction of variable stars in the instability strip is less than 1\%. 
This incidence of variable stars in NGC 2099 is much lower than the case of 
field stars with about 30\% \citep{bre79,wol83} and of some open clusters 
comparable to that of field stars \citep{hor79,fra96}. 
On the other hand, the low incidence of $\delta$ Scuti-type stars in 
NGC 2099 is in line with the cases of other many open clusters 
\citep[see][]{vis97}. We should note that our detection of variable stars near 
the $\delta$ Scuti instability strip are relied on the short exposure time 
data set with short time span, which are not enough to detect all 
multi-periodic small-amplitude $\delta$ Scuti-type stars in this cluster. 

Besides the KV1 and KV2, the other eight $\delta$ Scuti-type stars are 
deviated from the location of instability strip. They are fainter than the 
stars in the instability strip, indicating that they are not likely to be 
member stars in this cluster. According to their locations on the CMD, one 
peculiar variable KV10 and one RRc-type star V5 also are much less likely the 
member stars. Based on the analysis of empirical period-luminosity relation, 
\citet{kis01} also suggested that V5 and V6 are too faint to be bona-fide 
cluster member stars. 

As for the membership of eclipsing binary stars, we only consider three 
(V1, V2, and V4) binaries as possible members of the cluster, since they are 
brighter ($<$ 1 mag) than the main sequence for a given color, i.e. within 
the binary sequence. Another binaries fainter than the main-sequence should 
be field objects. 
However, two (KV12 and KV14) stars, which are slightly fainter than 
the main-sequence, can be marginal member stars considering their magnitude 
and color errors.

\section{Conclusions}

In the present study, we discovered nine $\delta$ Scuti-type stars, seven 
eclipsing binaries, and one peculiar variable in the field of open cluster 
NGC 2099. 
We also confirmed the identifications of five previously known variables 
by \citet{kis01}. While a number of stars are found in the $\delta$ Scuti 
instability strip of this cluster, we only detected two small amplitude 
$\delta$ Scuti-type stars. 
The other eight $\delta$ Scuti-type stars might be projected field stars, 
according to their locations in the CMD deviated to the $\delta$ Scuti 
instability strip. This is supported by the serious field star 
contamination due to the proximity of this cluster to the Galactic plane.  
Among seven newly discovered eclipsing binaries in our observations, six stars 
are W UMa-type binaries while one binary star is not clearly classified due to 
its incomplete phase diagram. 

KV10 shows peculiar light variations of $\delta$ Scuti-type short-periodic 
component combined with another long-periodic ones. 
The long-periodic variations can be interpreted by $\gamma$ Doradus-type 
pulsations or ellipsoidal variations. 
Because the former interpretation could not match with the infrared color 
indices of KV10, we prefer the possibility of ellipsoidal variations. 
Further spectroscopic or more precise multi-band photometric observations 
are needed to define physical characteristics of this interesting 
object. 

\acknowledgments
We thank Y.-J. Moon and Santabrata Das for their careful reading and 
comment. This research has made use of the WEBDA database, operated at the 
Institute for Astronomy of the University of Vienna. This study was 
financially supported by research fund of Chungnam National University in 2005.

\clearpage

\clearpage

\clearpage

\clearpage

\clearpage

\clearpage

\clearpage

\clearpage

\clearpage

\clearpage

\begin{deluxetable}{cccccc}
\tablecolumns{6}
\tablewidth{0pc}
\tablecaption{Observational summary of 24 variable stars in the field of 
NGC 2099.\label{tabl1}}
\tablehead{
\colhead{Star ID} &\colhead{RA.(J2000)} &\colhead{DEC.(J2000)} 
&\colhead{$<$V$>$} &\colhead{$<$\bv$>$} &\colhead{Type}
}
\startdata
V1   &05:52:20.354 &+32:33:20.42 &13.406 &0.609  &EA candidate\\
V2   &05:52:16.540 &+32:28:15.65 &14.653 &0.688  &EA candidate\\
V3   &05:52:33.022 &+32:32:41.79 &16.019 &0.514  &EW\\
V4   &05:52:53.272 &+32:33:01.33 &15.946 &0.851  &EW\\
V5   &05:53:00.644 &+32:24:50.59 &16.086 &0.811  &RRc\\
V6   &05:51:50.527 &+32:32:34.81 &16.177 &0.691  &$\delta$ Scuti\\
V7   &05:52:39.100 &+32:36:31.13 &17.623 &0.780  &EW\\
KV1  &05:52:34.319 &+32:32:18.74 &12.498 &0.427  &$\delta$ Scuti\\
KV2  &05:52:14.906 &+32:24:40.96 &13.298 &0.370  &$\delta$ Scuti candidate\\
KV3  &05:52:00.494 &+32:36:48.21 &14.762 &0.531  &$\delta$ Scuti\\
KV4  &05:53:10.442 &+32:33:47.30 &15.419 &0.390  &$\delta$ Scuti\\
KV5  &05:52:03.382 &+32:35:13.72 &16.039 &0.699  &$\delta$ Scuti\\
KV6  &05:52:44.697 &+32:30:16.52 &16.762 &0.368  &$\delta$ Scuti\\
KV7  &05:52:07.883 &+32:26:38.87 &16.780 &0.490  &$\delta$ Scuti\\
KV8  &05:52:11.171 &+32:25:15.81 &17.074 &0.700  &$\delta$ Scuti\\
KV9  &05:53:07.445 &+32:30:58.36 &17.865 &0.807  &$\delta$ Scuti candidate\\
KV10 &05:52:43.518 &+32:34:30.33 &17.325 &0.743  &$\delta$ Scuti+ellipsoid?\\
KV11 &05:52:10.999 &+32:41:45.20 &14.561 &0.450  &EW\\
KV12 &05:51:29.340 &+32:24:18.06 &16.406 &0.824  &E\\
KV13 &05:52:47.297 &+32:39:35.49 &17.648 &0.813  &EW\\
KV14 &05:52:40.708 &+32:24:24.35 &18.053 &1.078  &EW\\
KV15 &05:53:03.506 &+32:32:14.05 &18.540 &0.910  &EW\\
KV16 &05:52:43.889 &+32:28:52.22 &20.433 &0.865  &EW\\
KV17 &05:51:34.433 &+32:29:06.06 &20.751 &0.630  &EW\\
\enddata
\tablecomments{Star IDs of V are for the variables discovered by \citet{kis01} 
and those of KV are for the newly discovered variables by our observations.}
\end{deluxetable}

\clearpage

\begin{deluxetable}{cccccccc}
\tablecolumns{8}
\tablewidth{0pc}
\tablecaption{Variable stars identified by \citet{kis01}.\label{tabl2}}
\tablehead{
\colhead{}    &  \multicolumn{3}{c}{This study} &   \colhead{}   &
\multicolumn{3}{c}{\citet{kis01}} \\
\cline{2-4} \cline{6-8} \\
\colhead{Star ID} & \colhead{$\Delta$$V$(mag)}   & \colhead{Period(d)}    & \colhead{Epoch(H.J.D)} &
\colhead{}    & \colhead{$\Delta$$R_C$(mag)}   & \colhead{Period(d)}    & \colhead{Epoch(H.J.D)}}
\startdata
V1 &\nodata &\nodata &\nodata     &&0.24 &\nodata  &\nodata\\
V2 &\nodata &\nodata &\nodata     &&0.23 &\nodata  &2451540.5180\\
V3 &0.33    &0.4224  &2453014.660 &&0.31 &0.4224   &2451575.5083\\
V4 &0.32    &0.5582  &2453014.665 &&0.33 &0.5585   &2451576.7260\tablenotemark{\dag}\\
V5 &0.42    &0.2787  &2453014.634 &&0.32 &0.2800   &2451576.5000\\
V6 &0.52    &0.1098  &2453014.787 &&0.45 &0.1098   &2451540.5367\\
V7 &0.49    &0.3578  &2453014.792 &&0.55 &0.3579   &2451574.2620\\
\enddata
\tablenotetext{\dag}{Epoch shifted $+$0.5 phase from \citet{kis01}'s result 
which made a mis-identification of primary minimum due to the incomplete 
phase diagram.}
\end{deluxetable}

\clearpage

\begin{deluxetable}{ccccc}
\tablecolumns{5}
\tablewidth{0pc}
\tablecaption{Results of multiple frequency analysis for new pulsating 
variable stars.\label{tabl3}}
\tablehead{
\colhead{Star ID} &\colhead{Frequency(c/d)} &\colhead{$A_{j}$\tablenotemark{\dag}(mmag)} 
&\colhead{$\Phi_{j}$\tablenotemark{\dag}} &\colhead{S/N\tablenotemark{\ddag}}
}
\startdata
KV1  &$f_{1}$= 8.368 &11.3$\pm$.6  &3.14$\pm$.06  &8.5\\
KV2  &$f_{1}$=11.006 &7.7$\pm$.7   &$-$1.03$\pm$.10 &6.2\\
     &$f_{2}$=10.478 &4.5$\pm$.7   &2.10$\pm$.16  &4.9\\
KV3  &$f_{1}$=12.750 &16.1$\pm$.6  &$-$0.33$\pm$.03 &9.1\\
     &$f_{2}$=10.261 &10.9$\pm$.6  &$-$1.17$\pm$.05 &8.8\\
     &$f_{3}$=10.824 &10.3$\pm$.6  &2.19$\pm$.06  &9.7\\
     &$f_{4}$= 7.937 &6.6$\pm$.6   &1.21$\pm$.09  &7.6\\
     &$f_{5}$= 9.580 &4.7$\pm$.6   &$-$0.95$\pm$.13 &4.9\\
KV4  &$f_{1}$= 8.426 &52.1$\pm$.5  &0.54$\pm$.01  &34.2\\
     &$f_{2}$(=2$f_{4}$)=16.261 &12.0$\pm$.5  &4.70$\pm$.04  &11.2\\
     &$f_{3}$($\sim$$f_{1}$+$f_{4}$)=16.478 &8.3$\pm$.5   &$-$0.76$\pm$.06 &7.7\\
     &$f_{4}$= 8.129 &4.4$\pm$.5   &0.05$\pm$.11  &4.8\\
     &$f_{5}$=13.572 &3.9$\pm$.5   &3.93$\pm$.12  &4.3\\
     &$f_{6}$=15.856 &4.1$\pm$.5   &2.81$\pm$.12  &4.9\\
KV5  &$f_{1}$=12.719 &10.9$\pm$.5  &0.32$\pm$.05  &8.5\\
     &$f_{2}$= 7.889 &9.6$\pm$.5   &4.59$\pm$.05  &7.7\\
     &$f_{3}$=12.991 &8.1$\pm$.5   &2.31$\pm$.06  &8.6\\
     &$f_{4}$=14.450 &5.1$\pm$.5   &$-$0.02$\pm$.10 &6.2\\
     &$f_{5}$=11.018 &4.2$\pm$.5   &1.59$\pm$.12  &5.4\\
     &$f_{6}$= 9.511 &3.3$\pm$.5   &0.98$\pm$.15  &4.2\\
KV6  &$f_{1}$=10.037 &15.0$\pm$.6  &3.05$\pm$.04  &11.8\\
     &$f_{2}$=10.470 &8.7$\pm$.6   &$-$0.79$\pm$.06 &8.2\\
     &$f_{3}$=17.138 &5.2$\pm$.6   &4.68$\pm$.11  &5.8\\
     &$f_{4}$=14.034 &5.0$\pm$.6   &1.69$\pm$.11  &5.7\\
KV7  &$f_{1}$=24.308 &22.9$\pm$.6  &1.06$\pm$.03  &24.3\\
KV8  &$f_{1}$=15.173 &11.0$\pm$.8  &1.40$\pm$.07  &8.3\\
     &$f_{2}$=15.623 &11.2$\pm$.8  &2.31$\pm$.07  &8.9\\
     &$f_{3}$=16.227 &5.1$\pm$.8   &1.86$\pm$.15  &4.3\\
KV9  &$f_{1}$=13.020 &16.1$\pm$1.4 &3.91$\pm$.09  &8.0\\
\enddata
\tablenotetext{\dag}{$V=V_{0}+\Sigma_{j}A_{j}cos\{2\pi f_{j}(t-t_{0})+\Phi_{j}\}$, $t_{0} = H.J.D.2453000.0$}
\tablenotetext{\ddag}{S/N = (power for each frequency / mean power after 
prewhitening for all frequencies)$^{1/2}$}
\end{deluxetable}
\clearpage

\begin{deluxetable}{cccc}
\tablecolumns{4}
\tablewidth{0pc}
\tablecaption{Parameters of new eclipsing binaries.\label{tabl4}}
\tablehead{
\colhead{Star ID} &\colhead{Period(d)} &\colhead{Epoch(H.J.D)} &\colhead{$\Delta$$V$(mag)}}
\startdata
KV11 &1.887 &2453014.410 &0.15\\
KV12 &0.984 &2453014.470 &0.13\\
KV13 &0.289 &2453014.774 &0.19\\
KV14 &0.335 &2453029.635 &0.21\\
KV15 &0.355 &2453014.270 &0.23\\
KV16 &0.295 &2453013.815 &0.70\\
KV17 &0.224 &2453014.830 &0.75\\
\enddata
\end{deluxetable}

\end{document}